\documentclass[aps,prl,twocolumn,reprint,superscriptaddress, graphics,floatfix,tightenlines]{revtex4-2}
\usepackage[dvipsnames]{xcolor}
\usepackage{graphicx}
\usepackage{physics}
\usepackage{float}
\usepackage{caption}
\usepackage{hyperref}
\usepackage{mathtools}
\usepackage{breqn}
\usepackage{amsfonts}
\usepackage{dsfont}
\usepackage{amsmath}
\usepackage{mathtools}
\usepackage{xcolor}
\usepackage{bibunits}
\usepackage{subcaption}
\usepackage{soul}
\usepackage[nameinlink,capitalize]{cleveref}
\usepackage{amsmath}
\defaultbibliography{paper.bib}

\AtBeginDocument{\renewcommand*{\d}{\mathop{\kern0pt\mathrm{d}}\!{}}}
\defaultbibliographystyle{aipauth4-1}
\newcommand\myshade{85}
\colorlet{mylinkcolor}{violet}
\colorlet{mycitecolor}{YellowOrange}
\colorlet{myurlcolor}{Aquamarine}
\hypersetup{
  linkcolor  = mylinkcolor!\myshade!black,
  citecolor  = mycitecolor!\myshade!black,
  urlcolor   = myurlcolor!\myshade!black,
  colorlinks = true,
}
\begingroup\makeatletter
\catcode`,=\active
\global\let\breqn@comma,
\protected\gdef,{\ifmmode\expandafter\breqn@comma\else\expandafter\active@comma\fi}
\endgroup

\def\Arg{\mathop{\operator@font Arg}\nolimits}

\begin{document}
\title{Quantum constraint learning for quantum approximate optimization algorithm}
\author{\href{http://santoshkumarradha.me}{Santosh Kumar Radha}}
\email{Corresponding author:santosh@agnostiq.ai}
\affiliation{Agnostiq Inc., 180 Dundas Street W, Suite 2500, Toronto, ON M5G 1Z8, Canada}
\begin{bibunit}
\begin{abstract}
The quantum approximate optimization algorithm (QAOA) is a hybrid quantum-classical variational algorithm that offers the potential to handle combinatorial optimization problems. Introducing constraints in such combinatorial optimization problems poses a significant challenge in the extensions of QAOA to support relevant larger-scale problems. This paper introduces a quantum machine learning approach to learn the mixer Hamiltonian required to hard constrain the search subspace. We show that this method can be used for encoding any general form of constraints. One can directly plug the learnt unitary into the QAOA framework using an adaptable ansatz. This procedure gives the flexibility to control the depth of the circuit at the cost of the accuracy of enforcing the constraint, thus having immediate application in the Noisy Intermediate Scale Quantum (NISQ) era. We also develop an intuitive metric that uses Wasserstein distance to assess the performance of general approximate optimization algorithms with/without constraints. Finally, using this metric, we evaluate the performance of the proposed algorithm.
\end{abstract}
\maketitle
\paragraph{Introduction :} Variational quantum algorithms use short-depth circuits as a sub-routine in larger classical optimization and have been used in a wide range of contexts\cite{Kandala_2017,Havl_ek_2019,Schuld_2021,radha2021quantum}. The QAOA \cite{farhi2014quantum} is one such algorithm that has emerged as one of the most promising algorithms to use on NISQ computers for solving NP-hard combinatorial quadratic unconstrained binary optimization (QUBO) problems\cite{Lucas_2014}. More often than not, combinatorial optimization problems involve both equality and inequality constraints that need to be satisfied. 
\begin{figure}[t]
  \includegraphics[width=0.7\columnwidth]{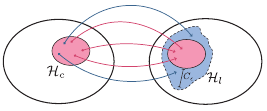}
  \caption{Total Hilbert space $\mathcal{H} $ $\supset$ constraint Hilbert space (pink) $\mathcal{H}_c$ being learnt with cost error $C_\epsilon$ leading to a leakage space $\mathcal{H}_l $ (blue) }
  \label{fig:hilbert}
\end{figure}
\begin{figure*}[t]
    \centering
    \begin{minipage}{0.49\textwidth}
        \centering
        \includegraphics[width=\columnwidth]{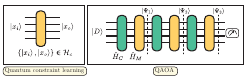} 
        \caption{QCL-QAOA \textit{(a)} First step -  learn constraint Hamiltonian $H_{qcl}$ by learning the propagator that propagates $\ket{x_i}\rightarrow\ket{x_o}$ $\forall \ket{x_{i,o}} \in \mathcal{H}_c$ \textit{(b)} Plugging the learnt $H_{qcl}$'s time evolved unitary inside the QAOA routine as a mixer. Each $\Psi_i$ leaks into unconstrained space with leakage proportional to the learning cost of $H_{qcl}$ }
        \label{fig:algo}
    \end{minipage}\hfill
    \begin{minipage}{0.49\textwidth}
        \centering
        \includegraphics[width=\columnwidth]{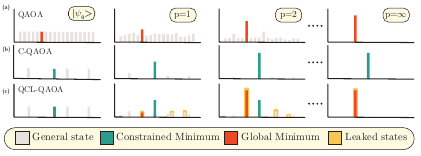} 
        \caption{Illustration of (a) QAOA (b) C-QAOA (c) QCL-QAOA 's optimized probability distribution  for various $p$ values and at $p\rightarrow \infty$. We start with initial state and as $p$ is increased, QAOA and C-QAOA converges to the global and constrained minimum respectively while QCL-QAOA has an optimal $p_0$ at which we maximize the constrain solution after which, due to leakage, it acts like general QAOA and converges to global optimum. }
        \label{fig:algo_compare}
    \end{minipage}
\end{figure*}

QAOA mainly consists of three main components - Initial state, objective function encoded in a diagonal Hamiltonian ($\hat{H}_{cost}$) and \textit{mixer} Hamiltonian ($\hat{H}_{mix}$) that mixes the states, with last two steps applied multiple times. Within this setting, constraints are applied in two major ways - soft and hard constrains\cite{Wang_2020,hadfield2018quantum}. Soft constrains\cite{fingerhuth2018quantum} are included in the initial optimization problems by adding a penalty term proportional to violating the constraint. This remaps the constrained search problem to that of an unconstrained problem and can be directly connected to the framework of QAOA. As a result, any general mixers like $X-$mixers can be used to solve the problem. Although soft constrain makes formalism easier, the cost landscape is plagued with complex local minimums, which are hard to optimize, and one is not guaranteed a solution that satisfies the constraint, hence the name soft constraint. In hard constrain, a more rigorous approach is followed by restricting the search space of the solutions to be inside the constrained space. Such \textit{hard} constraints are a subset of Quantum Alternating Operator Ansatz\cite{hadfield2018quantum,hadfield2019quantum}, where one designs a circuit in a way that the feasible expressibility of the ansatz is restricted to be in the search subspace. Thus, if one starts with a state in the constrained subspace (or a linear superposition of such), one is guaranteed to evolve it within the constrained space. Designing mixer Hamiltonians for restricting the subspace has been a significant challenge\cite{Hadfield_2019,Wang_2020,leipold2020constructing} with methods varying from mixing the feasible solutions using a $1D$ chain (\textit{ring mixers}) to mixing all pairs of feasible solutions (\textit{complete mixer}). These mixers are designed for a specific class of constraints, not for general constraints. Therefore, finding a method to encode general constraints has a considerable value.Meanwhile, using quantum circuits for machine learning tasks has gathered a lot of attention\cite{wittek2014quantum,biamonte2017quantum,schuld2018supervised}. The use of specifically designed circuits and the corresponding cost function defines the application. For instance, QAOA can be thought of as a circuit learning system for ground-state preparation, with the cost function being the expectation value of $\hat{H}_{cost}$.

In this letter, we solve the problem of designing the mixer Hamiltonian by leveraging quantum machine learning to \textit{learn} the mixer Hamiltonian in order to enforce the hard constrained search space. In a nutshell, given a $\log_2{n}$ dimensional cost problem in $n$ dimensional Hilbert space $\mathcal{H}$ along with constraints which belong to the subspace $\mathcal{H}_c \subset \mathcal{H}$, our goal is to learn the Quantum Constraint Learnt (QCL) Hamiltonian $H_{qcl}$, the unitary time evolution of which enables us to mix and restrict a vector within the constrained subspace. More often than not, in real-world use cases, the constraints are fixed and seldom change. For example, in the case of portfolio optimization problems, the weighted sum of qubits must always be $100\%$\cite{hodson2019portfolio}. Thus such a technique that requires an initial overhead of learning the Hamiltonian that reduces the number of iteration per task can be advantageous if the same task is performed routinely and repetitively. As the problem of designing the mixer Hamiltonian is translated into the problem of learning it, one can tune trade-off between circuit depth and the accuracy of approximating exact mixers.

\paragraph{Constraint learning :} There are two major criteria that $\hat{H}_{qcl}$ needs to satisfy. 
\begin{align}
e^{-i H_{qcl} t}\ket{\psi_i} &\in  \mathcal{H}_c  \label{eq:c1}\\
\bra{\psi_j}e^{-i H_{qcl} t}\ket{\psi_i}&>0\label{eq:c2}
\end{align}

$\forall t\in \mathbb{R}$ and $\forall \{\ket{\psi_i},\ket{\psi_j} \}\in  \mathcal{H}_c$. \autoref{eq:c1} ensures any time evolution of any state vector in $\mathcal{H}_c$ does not leave $\mathcal{H}_c$. This is illustrated schematically in \autoref{fig:hilbert} where the entire black circle represents $\mathcal{H} $ and the pink subspace, $\mathcal{H}_c$. The action of $e^{-i H_{qcl} t}$ maps the pink area back to itself. \autoref{eq:c2} ensures that there exists transitions between all possible vectors inside the constrained space. Since we are working with $\hat{H}_{cost}$ which is diagonal in the computational basis, one can restrict $\ket{\psi_i}$ to just vectors in the computational basis that satisfy the constraint \textit{i.e.} binary bit strings that satisfy the constraint. The main approach we take is to formulate the problem of finding such $\hat{H}_{qcl}$ as a variational optimization problem. To this end, we take inspiration from variational diagonalization method proposed in Ref\cite{larose2019variational,cirstoiu2020variational} to guide the design of our variational ansatz. We use the circuit of the form $V(\boldsymbol{\alpha},t)$ with
\begin{equation}
  V_{lk}(\boldsymbol{\alpha}, t):=W_l(\boldsymbol{\theta}) D_k(\boldsymbol{\gamma}, t) W_l(\boldsymbol{\theta})^{\dagger} =e^{-i H_{qcl}(\boldsymbol{\theta},\boldsymbol{\gamma}) t}\label{eq:1},
\end{equation}
where $\boldsymbol{\alpha}=\{\boldsymbol{\gamma}, \boldsymbol{\theta}\}$ and $W_l(\boldsymbol{\theta})$ is a $l$ layered parameterized quantum circuit (PQC) , $D_k(\boldsymbol{\gamma}, t)$ is a $k-$local parameterized diagonal unitary evolved for time $t$\cite{sm}. Based on the criteria that $H_{qcl}$ needs to satisfy, we develop two variations of cost function to be minimized. Without loss of generality, for a given $n$ dimensional QUBO problem, we shall assume that the constraint space is spanned by the binary bit string vectors $\ket{x_i}$ with $i \in \{0,1,\ldots m-1\}\quad m<2^n-1$.

\paragraph*{One-to-one}: Very intuitive variational cost function is to minimize 
\begin{align}
C_1(\boldsymbol{\alpha})=\int dt \sum_{ij}\matrixel{x_i}{V_{lk}(\boldsymbol{\alpha}, t)}{x_j}.
\end{align}

Pairs of $(i,j)$ that goes into the summation determines the connectivity of the resultant mixer Hamiltonian's graph. Based on the appetite of mixing one can choose the exhaustiveness of $\ket{x}$'s in the sum making up $C_1$. For instance in the case of ring mixers, one can just chose the $m$ set of pairs $\{(\ket{x_0},\ket{x_1})\ldots ,(\ket{x_i},\ket{x_{i+1}})\ldots ,(\ket{x_m},\ket{x_{0}})\}$ to sum up, while for a complete graph, there are in total $m(m-1)/2$ terms in the summation.

\paragraph*{One-to-many}: Instead of learning to map each individual constrained state to another, one can map the constrained state $\ket{x_j} \rightarrow e^{-i\theta_j} m^{-1/2} \sum_{i=0}^{m-1} \ket{x_i}$. This maps every constrained state to a superposition of all constrained states. Because of the unitary nature of mapping, one cannot have two states mapping to the same resultant state, phase factor $e^{-i\theta_j}$ prevents it from happening. To construct a cost function to do the same, we denote  $V_{lk}(\boldsymbol{\alpha}, t)\ket{x_i}=\sum_{j=0}^{2^n-1} \sqrt{p_j^i(\boldsymbol{\alpha},t)}\ket{j}$ and let $\boldsymbol{p}^i(\boldsymbol{\alpha},t)=[p^i_0(\boldsymbol{\alpha},t),\ldots,p^i_{2^n-1}(\boldsymbol{\alpha},t)]$ be the probability vector for a given input state $\ket{x_i}$. We denote $\tilde{\boldsymbol{p}}$  as the probability vector for a uniformly superimposed constrain states. With this, one can now write the variational cost to minimize -
\begin{align}
  C_2(\boldsymbol{\alpha})=\int dt \sum_i D(\boldsymbol{p}^i(\boldsymbol{\alpha},t),\tilde{\boldsymbol{p}}),
\end{align}

where $D(\boldsymbol{p},\boldsymbol{q})=\sum_i p_i \log{\frac{p_i}{q_i}}$.

\paragraph*{Algorithm}: First, we start with $l-$layered, $k-$local PQC $V_{lk}(\boldsymbol{\alpha})$. We then find the optimal parameter set $\boldsymbol{\alpha}^*$ by minimizing either of the two cost functions ($C_1$,$C_2$), which can be efficiently computed in a quantum computer. This is the constraint learning part that finds the unitary $U(t_i)=V(\boldsymbol{\alpha}^*,t_i)=\exp{-i\hat{H}_{qcl}t_i}$ as shown in \autoref{fig:algo}(a). For all practical purposes, choosing $k=1$ and converting the integral over time to a single time step (for numerical results in the paper, we use $t_0=1$) turns out to be a good enough choice. Independent of the given QUBO problem, this first part needs to be done only once for a given set of constraints. Next, following the standard procedure of QAOA, we then encode the optimization problem as Hamiltonian $\hat{H}_c$ and start with an initial state $\ket{D}$ which could either be a simple binary bit inside the constraint space or a linear superposition of those. For a $p$ depth QAOA,  we then apply the unitary $\exp{-i\gamma \hat{H}_c} \exp{-i\hat{H}_{qcl}\beta}$ exactly $p$ times to the initial state and make a final measurement of the expectation value of $\hat{H}_c$ which is then minimized in a hybrid classical-quantum fashion by optimizing $2p$ variables $\{\boldsymbol{\gamma},\boldsymbol{\beta} \}$.

It is important to note that we are not variationally learning multiple independent unitaries $V_{lk}(\boldsymbol{\alpha}, t):=U_t$ for various $t's$ but rather we are learning the strongly continuous one-parameter unitary group $U_t$ for which there always exists a unique (possibly unbounded) operator $H$ which is the infinitesimal generator of $U_t$, such that $\forall t \in \mathbb{R}$, $U_t=e^{-iHt}$. $H$ is guaranteed to exist and is unique by Stone's theorem\cite{stone1932one}. Thus, we are learning this Hamiltonian $H$.

\begin{figure*}[!htb]
  \includegraphics[width=\linewidth]{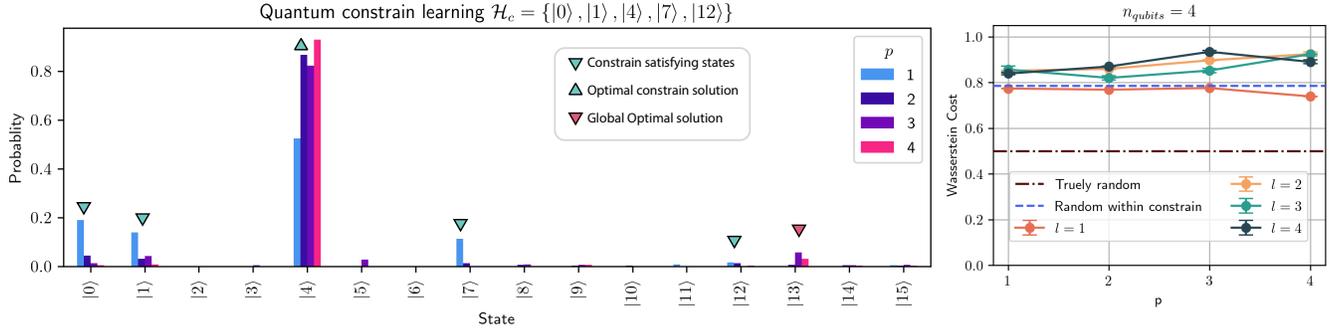}
  \caption{\textit{(a)} Probability distribution of optimized QCL-QAOA for a random instance of QUBO problem with $\ket{13}$ being the global minimum and $\ket{4}$ being the constrained minimum for various values of $p$. Downward pointing green triangles denote the solutions that satisfy the constrain while the upward green pointing one is the ideal constrained optimum. Red triangle is the unconstrained global minimum solution. \textit{(b)} QCL-QAOA performance as function of depth $p$ for various values of layer depth $l$ }
  \label{fig:algo_result}
\end{figure*}

\paragraph{Discussion :}The mixer Hamiltonian can be viewed in multiple ways. First, mixing the unitary of the Hamiltonian is nothing but a propagator $\mathcal{K}(x_i,x_j,t)$ with the property that there exists at least one time $t$ where the propagating kernel connecting the states $x_i,x_j$ within the constrained space is non-zero. Second, as pointed out by Ref\cite{marsh2019quantum}, one could think of mixer as a random walk of wave functions exploring the solution subspace and thus construct the mixer based on the corresponding random walk's graph. The mixing power of the mixer can be directly correlated with the connectedness of the graph, which implies that a well-connected constrained mixer graph leads to a better solution. Connections of the constrained graph directly influence the complexity of the learning process in the case of $C_1$. The number of terms in the cost function is independent of QUBO dimension but dependent on the number of constraints with the terms scaling at the best $O(m)$ for ring mixers and the worst $O(m^2)$ for complete mixer.

Before diving into the algorithm's performance, we qualitatively discuss the subtleties of QCL compared to QAOA and other methods of encoding constraints. In a standard QAOA setup, one starts with an equal superposition of all solutions, and as deeper $(p)$ ansatz is added, one expects the solution to get closer and closer to the optimal one. In the case of constrained QAOA (C-QAOA), one starts with an initial state that is either a superposition of all constrained solutions or one of the constrained solutions, and as $p$ increases, one gets closer to the constrained optimal solution. In the case of QCL-QAOA, one encounters a non-trivial $p$ dependence for the quality of the solution. This is shown illustratively in \autoref{fig:algo_compare}, (a) shows the general QAOA, (b) C-QAOA and (c) QCL-QAOA. Because of the errors in learning the mixer Hamiltonian, which is precisely given by $C_{(1,2)}\neq 0$, non-zero connections form between the constrained subspace and the unconstrained one (shown as blue lines in \autoref{fig:hilbert}). These connections, in turn, leak the wave function into the unconstrained search space. We dub this phenomena \textit{quantum leakage}. As a result of quantum leakage, when applying multiple mixer unitaries (increasing $p$), we end up leaking the wave function to unconstrained outer space. As one increases $p$, QCL-QAOA is bound to find the global minimum of the search space and not the constrained minimum. Thus, there is an optimal depth $p_0$ after which the quality of the solution decreases. In principle, one expects $p_0$ to be function of the error $C_{(1,2)}$.

To illustrate the algorithm in action, we choose the cost function $C_1$ and a complete mixer graph to represent the constraint for our problem size of $4$ qubits. Since we are learning the mixer Hamiltonian, we can encode arbitrary constrains, and we chose the states $\ket{i}$ where $i\in\{0,1,4,7,12\}$ as constrain satisfying states and train the Hamiltonian. We make sure to design the QUBO problem such that the global minimum is outside the constrained search space. Note that this makes it harder to solve the problem than when it is inside the constrained search space, which is solvable by a standard QAOA. We then use a combination of grid search and COBYLA\cite{powell1998direct} as a method of optimization for each CQL-QAOA instance. In \autoref{fig:algo_result}(a), we show one such instance of the constraints being the states $\ket{i}$ where $i\in\{0,1,4,7,12\}$ with states $\ket{13}$ ,$\ket{4}$, the global and constrained minimum respectively. With this learnt mixer, we then proceed with QCL-QAOA at various depths, the results of which are plotted in (a). First, the search is almost constrained to be within the subspace. Second, for higher $p>1$, despite the improved quality of solution, we see probability creeping into the unconstrained space ($\ket{13}$). As discussed above, this illustrates the competing behavior of QAOA and quantum leakage trying to improve and impair the solution respectively as a function of depth ($p$).

\paragraph{Performance metric:} To better understand the non-trivial $p$ dependence and to asses the performance of QCL-QAOA, one needs to quantify the quality of the results. Ref.\cite{farhi2014quantum} introduced approximation ratio $r$ as a performance metric, which is given by
\begin{align}
  r=\frac{\bra{\psi^*}\hat{H}_{cost}\ket{\psi^*}}{\tilde{C}} \label{eq:4},
\end{align}
 where $\ket{\psi^*}$ is the optimal parameterized wave function and $\tilde{C}$ is the optimal value of the given QUBO problem. First, the numerator of \autoref{eq:4} could, in theory, have a different sign to that of $\tilde{C}$ leading to negative $r$. Second, the meaning of $r$ is lost in the case of constrained QUBO problems. Even if one potentially modifies the $\tilde{C}$ to be the constrained optimum, the value would be unbounded when the algorithm leaks solution into the unconstrained space. To this end, we construct a metric that considers these factors. We propose the use of Wasserstein distance\cite{villani2003topics}, which, for two probability distributions $p,q$ in $\mathbb{R}$ with cumulative distribution $P,Q$ respectively, is given by 

 \begin{align}
  W(p, q)=\int_{\mathbb{R}}\abs{P(y)-Q(y)}dy \label{eq:5},
 \end{align}
 
 as a measure of performance. To do so, we construct a measurable space, and the inherent cost associated with each $\ket{i}$ already forms a measurable space. However, to distinguish the performance within the constrained space, we reorder the states by their ability to satisfy the constraint first and then by the ascending order of their cost. To have the performance metric bounded, we define the distance between each adjacent state to be $\frac{1}{2^n}$ instead of their corresponding cost. We call this new ordered basis $\ket{\tilde{i}}$. It is trivial to see that the $W\left(\ket{\tilde{0}},\ket{\tilde{\psi}}\right)$ is an exact measure of how close we are to the solution\cite{sm}. To be consistent with $r$, we instead choose $\zeta\left(\ket{\psi}\right)=1-W\left(\ket{\tilde{0}},\ket{\psi}\right)$. We thus have $ \zeta \in [0,1]$ with $0$ being the worst performer and $1$ being the best. Notice that a uniform superposition of all states (classical random algorithm) gives a performance value of $0.5$. $\zeta$ indicates both the quality of solution (how concentrated/spread out the probability of solution is) as well as the value of the solution (at which cost the solution has its peak).

In \autoref{fig:algo_result}(b), we show $\zeta$ for various values of $l$ and $p$. One achieves higher expressivity of the mixer's PQC as $l$ increases, this, in turn, reduces the corresponding optimal cost $C_1(l)$. Each data point is a collection of $50$ instances of QUBO problem solved using QCL-QAOA. We also mark the \textit{truly random algorithm} (equal superposition of all states) and \textit{random within constraints} algorithm (equal superposition of constrained states) by black and red, respectively. One can think of \textit{random within constraints} algorithm as the basic metric that a hard constrained search needs to be greater than, as any leakage would most probably have a value smaller than that. Because of the inefficiency to learn the mixer, at $l=1$, we see that the algorithm barely constrains the search space. As $p$ is increased, due to the high cost value (hence high quantum leakage), the performance counter intuitively decreases drastically. By contrast, for $l\in[2,3,4]$, we see that the performance increases as $p$ is increased as one would expect. For $l=4$, we see the formation of a peak at $p_0=3$, after which the quantum leakage takes over, and the performances decreases. Overall, we see that QCL-QAOA has restricted the search space and produced a high-quality solution.

\paragraph{Summary:} To summarize, we propose an algorithm to inject arbitrary constraints in QAOA. The algorithm uses quantum machine learning to learn the constrained Hamiltonian using an adaptable ansatz. After learning, the Hamiltonian is fed into the standard QAOA framework as a mixer unitary. This method of implementing constraints in QAOA takes away the need to design complex mixer Hamiltonians and can also be used to encode arbitrary constraints. We finally assess the proposed algorithm's performance using a metric explicitly designed for constrained QUBO problems. This metric, at its core, uses Wasserstein distance and intuitively describes the quality of the solution. A major advantage of this method is the flexibility to choose the depth of the learnt mixer unitary which in turn translates to the depth of the final QAOA, albeit at the cost of performance, which is of huge importance in the NISQ era. Finally, we would like to note that due to the nature of the ansatz, one can directly deduce the learnt mixer Hamiltonian, which can then be used to constrain quantum annealers\cite{PhysRevA.93.062312}. Recently, the problem of finding the constrained driver Hamiltonian was shown to be a NP-complete problem\cite{leipold2020constructing}.
\putbib

\end{bibunit}

\begin{bibunit}

\clearpage
\pagebreak
\setcounter{equation}{0}
\setcounter{figure}{0}
\setcounter{table}{0}
\setcounter{page}{1}
\makeatletter
\renewcommand{\theequation}{S\arabic{equation}}
\renewcommand{\thefigure}{S\arabic{figure}}
\renewcommand{\bibnumfmt}[1]{[S#1]}
\renewcommand{\citenumfont}[1]{S#1}

\widetext
\begin{center}
\textbf{\large Supplemental Material: Quantum constraint learning for quantum approximate optimization algorithm}
\end{center}
Supplemental material contains various details that extend the ideas presented in the main paper. We briefly overview the Quantum Approximate Optimization Algorithm and its generalized counterpart Quantum alternating operator ansatz. We then discuss the circuits used to compute the two cost functions introduced in the main paper and comment on the cost landscape of this learning routine. We discuss the quantum leakage phenomena with numerical examples quantifying it and providing more intuition on how various parameters affect it. Finally, we explain the Wasserstein performance metric discussed in the main paper.

\section{Quantum Approximate Optimization Algorithm}\label{sec:sm-qaoa1}

Quantum Approximate Optimization Algorithm (QAOA)\cite{farhi2014quantum}, and the extension of its framework to the much more general class of quantum alternating operator ansatz is quantum heuristics for approximately solving classical optimization problems which are NP-hard. Low-depth versions of these algorithms are suitable for NISQ devices as they entail a hybrid classical-quantum routine. Parameterized quantum circuits are used for efficiently calculating an expectation value (cost function), which are then updated with the help of a classical loop. While in some cases, suitable parameters can be found analytically\cite{PhysRevA.95.062317} or through decomposition techniques\cite{Streif_2020}, variational approaches remain the primary choice. Although experimental assessments of QAOA have been performed \cite{Harrigan_2021}, very little is known about the performance and its resilience to noise\cite{Zhou_2020,marshall2020characterizing,hastings2019classical}. 

 Quadratic unconstrained binary optimization (QUBO) problem is given by $\min\limits_{x\in \{0,1\}}\boldsymbol{x}\mathcal{Q}\boldsymbol{x}^T$, where $\mathcal{Q}$ is an upper diagonal matrix. To start with, we rewrite the QUBO cost function to that of a classical spin Ising Hamiltonian, which we will call $H_{cost}$. This is done in such a way that the eigen-solutions of the Hamiltonian are directly related to the classical QUBO's cost problem. The algorithm consists of two unitaries made up of the cost Hamiltonian $H_{cost}$ and mixing Hamiltonian $H_{mixer}$. As described, the cost Hamiltonian encodes the cost and acts diagonally on the given $n$ qubit system as 

\begin{align}
  H_{cost}\ket{x}=c(x)\ket{x}\label{eq:s1},
\end{align}
where $c(x)$ is the QUBO cost corresponding to the bit string $x$. The mixing unitary, usually taken to be $X-$mixer is the transverse field Hamiltonian
\begin{align}
  H_{mixer}=\sum_{i=1}^{n}X_i,\label{eq:s2}
\end{align}
where $X_i$ is the Pauli $X$ operator that acts only on $i^{th}$ qubit. Note that this operator acts as a bitflip operator \textit{i.e.} $X\ket{0}=\ket{1}$ and $X\ket{1}=\ket{0}$. With these two basic ingredients, one starts with an initial state $\ket{\psi_i}$, usually taken to be a superposition of all possible solutions, which is given by
\begin{align}
  \ket{\psi_i}=\sum_x \sqrt{2^{-n}}\ket{x}\label{sq:s3},
\end{align}
and creates a variational state that depends on $2p$ parameters $(\boldsymbol{\beta}, \boldsymbol{\gamma})$ by alternatively applying the evolution of $H_{cost}$ and $H_{mixer}$. This results in the following variational state 
\begin{align}
  |\boldsymbol{\beta}, \boldsymbol{\gamma}\rangle=e^{-i \beta_{p} H_{mixer}} e^{-i \gamma_{p} H_{cost}} \ldots e^{-i \beta_{2} H_{mixer}} e^{-i \gamma_{2} H_{cost}} e^{-i \beta_{1} H_{mixer}} e^{-i \gamma_{1} H_{cost}}|\psi_i\rangle.
\end{align}

Finally, measurements are made on the resulting state $|\boldsymbol{\beta}, \boldsymbol{\gamma}\rangle$ which gives us solution $x$ with probability $\abs{\braket{x}{\boldsymbol{\beta}, \boldsymbol{\gamma}}}^2$; using which one calculates the expectation value of $H_{cost}$, given by
\begin{align}
  f(\boldsymbol{\beta}, \boldsymbol{\gamma})=\expectationvalue{H_{cost}}{\boldsymbol{\beta}, \boldsymbol{\gamma}}\label{eq:s5}.
\end{align}

We then uses a classical loop to optimize $f(\boldsymbol{\beta}, \boldsymbol{\gamma})$  over the continuous $2p$ parameters using techniques like gradient descent. In the ideal case, the final optimized wave function solution should have its probability concentrated on the solution bit string $\ket{x^*}$ 

\subsection{Quantum alternating operator ansatz}\label{sec:sm-qaoa2}
Quantum alternating operator ansatz\cite{Gard_2020,PhysRevA.101.052340,Ryabinkin_2018}, is an extension of QAOA, where one chooses two general unitaries - $U_c$, which depends on the objective function and $U_m$, that depends on the structure of the solution (like constraints)  that are applied in an alternating fashion to output the desired final state. These unitaries responsible for \textit{phase separation} and \textit{mixing} the initial wave function $\ket{\psi_i}$ respectively. It is trivial to see that QAOA is a subclass of quantum alternating operator ansatz with $e^{-i \beta_{p} H_{mixer}} , e^{-i \gamma_{p} H_{cost}}$ playing the roles of unitary, each of which is responsible for evolving the initial wave function for time $\beta_p$ and $\gamma_p$ respectively. It is important to note that often, this extension is used to encode hard constraints by choosing $U_c$ to be the same as that of QAOA and $U_m$ to encode the restricted search space, for instance, in Ref.\cite{Wang_2020} $XY$ Pauli matrices are used as mixers which are designed to restrict the search Hilbert space. We use quantum machine learning to learn the mixer Hamiltonian which restricts the subspace.

\section{Quantum Constraint learning}\label{sec:sm-qcl}
\begin{figure*}
  \includegraphics[width=0.8\linewidth]{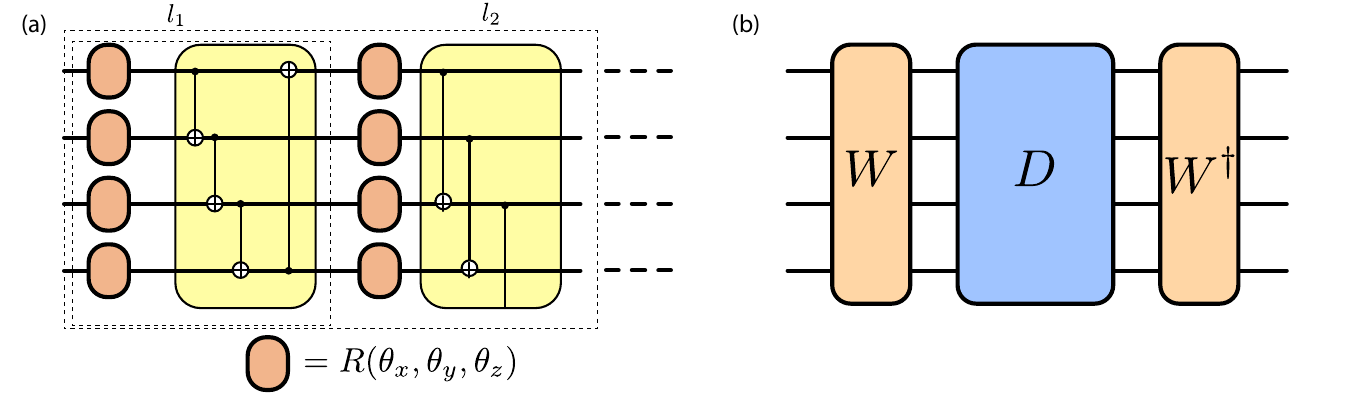}
  \caption{(a) PQC for variational ansatz $V$, each orange block represents an instance of $R(\theta_x,\theta_y,\theta_z)$ (b) Circuit representation of the $\exp{-i\beta H_{mixer}}$}
  \label{fig:sm_pqc_learning}
\end{figure*}

This section describes the two quantum machine learning techniques used in the main text in detail. For learning the mixer Hamiltonian developed in Quantum Constraint Learning (QCL), we use the ansatz developed in Ref\cite{larose2019variational,cirstoiu2020variational}. The consideration for developing a learning ansatz for the mixer Hamiltonian is to encode the required transition between the constrained space and the ease of access to time propagation of the corresponding unitary of the Hamiltonian. This is done by choosing a variational form given by 
\begin{equation}
  V_{lk}(\boldsymbol{\alpha}, t):=W_l(\boldsymbol{\theta}) D_k(\boldsymbol{\gamma}, t) W_l(\boldsymbol{\theta})^{\dagger} =e^{-i H_{qcl}(\boldsymbol{\theta},\boldsymbol{\gamma}) t} \label{eq:sm6},
\end{equation}
where $W_l(\boldsymbol{\theta})$ is a $l$ layered parameterized quantum circuit (PQC) , $D_k(\boldsymbol{\gamma}, t)$ is a $k-$local parameterized diagonal unitary evolved for time $t$and $\boldsymbol{\alpha}=\{\boldsymbol{\gamma}, \boldsymbol{\theta}\}$. Thus, the action of $W$ here could be regarded as a basis change to the eigen-basis where the required Hamiltonian is diagonal. In this new basis, ($t$) time evolution of the initial state with respect to this Hamiltonian, is nothing but $\exp{-it D_{jj}}$, where $D_{jj}$ are the eigenvalues of the Hamiltonian. Once the state is evolved for the said time, it is converted back to its initial basis by the application of $W^{\dagger}$. Thus, with this ansatz, to evolve the given state by time $\beta_i$, all one needs to do is act the unitary $V(\boldsymbol{\alpha}, \beta_it)$ to the initial state. This circuit is shown in \autoref{fig:sm_pqc_learning}(b).

Although one could in principle choose any ansatz for $W$ including hardware-efficient ansatz\cite{Kandala_2017}, low-depth ansatz\cite{PhysRevApplied.11.044087}, tensor-network ansatz \cite{Huggins_2019} etc., here, since we do not focus on the efficiencies of the ansatz, we choose an ansatz inspired from the \textit{Strongly entangling circuits} as introduced in Ref.\cite{PhysRevA.101.032308}. 

In \autoref{fig:sm_pqc_learning}(a) we show the modified PQC we use for generating the data in this letter. Each layer of a $l$ layer PQC consists of single layer of $x,y,z$ rotation gates for all qubits, followed by entangling layers made up of $CNOT$ gates. The entangling gates at $l_i^{th}$ layer consists of $CNOT$ gates entangling qubits $j$ and $(j+i)\%n$ $\forall j \in [1,n]$. 

As noted by Ref.\cite{cirstoiu2020variational}, constructing a diagonal unitary $D$ is equivalent to finding a Walsh series approximation
\begin{align}
D=e^{i G}=\prod_{j=0}^{2^{q}-1} e^{i \gamma_{j}} \otimes_{k=1}^{n}\left(Z_{k}\right)^{j_{k}} \label{eq:smdiag}
\end{align}
where $q=n$ , $G$ and $Z_k$ are diagonal operators with pauli operator $Z_k$ acting on $k-^{th}$ bit in bitstring $j$. Efficient quantum circuits for minimum depth approximations of $D$  is obtained by resampling the function on the diagonal of $G$ at sequences lower than a fixed threshold, with $q=k$, with $k\leq n$. The resampled diagonal takes the same form as \autoref{eq:smdiag} but with $q=k$\cite{welch2014efficient}. Unlike Ref\cite{cirstoiu2020variational}, since we are not dealing with physical Hamiltonians, there is no correlation between the locality of the diagonal terms and the system. We can thus choose an arbitrary set of terms in the $2^q-1$ terms for the approximation based on hardware constraints.

In the main paper, we discussed two kinds of cost functions that can be used to learn the Hamiltonian. The first one is the \textit{One to one} while the second one, \textit{One to many} has much lower QPU calls compared to the first. In the following subsections, we describe these cost functions and the circuit used to calculate them. Through the next subsections, we assume we have a $n$ dimensional QUBO problem, with constraint space spanned by the binary bit string vectors $\ket{x_i}$ with $i \in \{0,1,\ldots m-1\}\quad m<2^n-1$.

\begin{figure*}
  \includegraphics[width=0.9\linewidth]{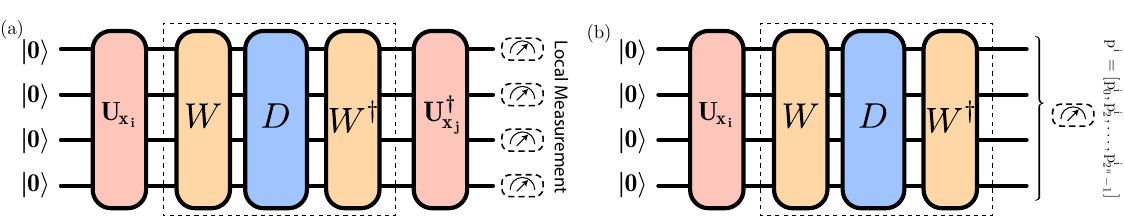}
  \caption{(a) Circuit for calculating terms $\matrixel{x_i}{V_{lk}(\boldsymbol{\alpha}, t)}{x_j}$ in one-to-one cost function $C_1$ given in \autoref{eq:sm7}  (b) Circuit for calculating the probability $\boldsymbol{p}^i$ for one-to-many cost function $C_2$}
  \label{fig:costcircuit}
\end{figure*}

\subsection{One to one}

As discussed in the main text, the cost function of one-to-one circuit is given by
\begin{align}
C_1(\boldsymbol{\alpha})&=\int dt \sum_{ij}\matrixel{x_i}{V_{lk}(\boldsymbol{\alpha}, t)}{x_j}\label{eq:sm7},\\
&\approx \sum_{ijk} \matrixel{x_i}{V_{lk}(\boldsymbol{\alpha}, t_k)}{x_j}\label{eq:sm8},
\end{align}

where $\ket{x_i}$ and $\ket{x_j}$ belong to the constrained search space. \autoref{eq:sm7}`s integral of $t$ is turned into summation over a finite set of $t$'s $\{t_1,t_2\ldots t_k\}$. Throughout our experiments, for up to $8$ qubit system setting $k=1$ and $t_1=1$ seems to be sufficient. The circuit for calculating the terms in summation is given in \autoref{fig:costcircuit}(a). First, a binary encoding unitary $U_{x_i}$ is applied, which initializes the binary state $\ket{x_i}$ after which our ansatz $V=WDW^\dagger$ is applied followed by $U_{x_j}^\dagger$. For optimal ansatz, $VU_{x_i}\ket{0^{\otimes n}}=\ket{x_j}$ and thus when final unitary $U_{x_j}^\dagger$ is applied, the resulting state is $\ket{0^{\otimes n}}$. We then use the local cost function where we measure and calculate the mean probability $(P_0)$ of getting $\ket{0}$ in all individual qubits, which is then maximized. This computes the local cost of the parameterized mixer Hamiltonian, taking the state $\ket{x_i}$  to $\ket{x_j}$ after which one sum over all such $i,j$ within the constraint graph. Note that in cases where $i$ maps to multiple $j$ (complete mixer), the total cost can never get to $0$. However, minimizing the cost will result in $i$ mapping to a superposition of these states. 

\subsection{One to Many}

Like the one-to-one cost function, we start by discretizing the time integral and choosing a single time step for all numerical results with $t=1$. Thus we get 

\begin{align}
  C_2(\boldsymbol{\alpha})= \sum_{i,t=1} D(\boldsymbol{p}^i(\boldsymbol{\alpha},t),\tilde{\boldsymbol{p}}),\label{eq:smo9}
\end{align}

where $\boldsymbol{p}^i(\boldsymbol{\alpha},t)=[p^i_0(\boldsymbol{\alpha},t),\ldots,p^i_{2^n-1}(\boldsymbol{\alpha},t)]$ is the measured probability vector as given in \autoref{fig:costcircuit}(b) and $\tilde{\boldsymbol{p}}=[\tilde{p}^i_0,\tilde{p}^i_1\ldots,\tilde{p}^i_{2^n-1}]$  is given by

\begin{align}
\tilde{p_j}=\left\{ 
  \begin{array}{ c l }
    m^{-1} & \quad \textrm{if } \ket{j} \in \mathcal{H}_c \\
    0                 & \quad \textrm{otherwise}
  \end{array}
\right.,
\end{align}
which is the probability vector for uniformly super imposed states that belong to the constrained space. $D(\cdot,\cdot)$ could be any measure of distance between probability distributions including Kullback-Leibler (KL) divergence, Euclidian norm, wasserstein distance etc. For numerical results in this paper, we use $D_{KL}(\boldsymbol{p},\boldsymbol{q})=\sum_i p_i \log{\frac{p_i}{q_i}}$.

\subsection{Cost landscape discussion} 
Barren plateau is a phenomenon in which, for a broad class of reasonable parameterized quantum circuits, the probability that the gradient along any reasonable direction is non-zero to some fixed precision is exponentially small as a function of the number of qubits\cite{McClean_2018}. It was shown that a cost function with local observables leads to, at worst, a polynomially vanishing gradient\cite{Cerezo_2021}. Both cost functions introduced above are local and, hence, in their purest form, do not have barren plateaus issues. Although both these costs are local, in our numerical simulations, we observe that in the case of complete mixers, the one-to-many cost function converges much faster than the one-to-one. This is most probably an outcome of the inherent one-to-many input-output mapping of the cost function, which leads to the optimization getting stuck between the multiple local minima or saddle points. This again was verified by having a one-to-one mapping (ring mixer) where both the cost functions converge to similar values with the same resource.

\section{Quantum leakage}
\begin{figure}[h]
  \includegraphics[width=0.6\columnwidth]{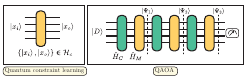}
  \caption{QCL-QAOA \textit{(a)} Learning the constrained unitary \textit{(b)} Plugging the constrained unitary inside the QAOA routine as a mixer, $\ket{\psi_i}$ denote the state of wave function at each layer of the circuit.}
  \label{fig:sm_algo}
\end{figure}

In the main paper, we briefly outlined the phenomena of quantum leakage. This section will provide a more detailed discussion on the topic. In \autoref{fig:sm_algo}(b), we show the QCL-QAOA procedure where the learnt mixer unitary is fitted into the QAOA routine. The role of the constrained mixer is to create connections/ interactions between the constrained subspace that allows for mixing and exploring all solutions (or most of them) inside the feasible subspace. In the presence of learning error, the mixer instead creates interactions outside the constrained space. This is what we dub quantum leakage, where an initial state starting inside the constrained space can now \textit{leak} into the unconstrained space.

\begin{figure}[ht]
  \includegraphics[width=\columnwidth]{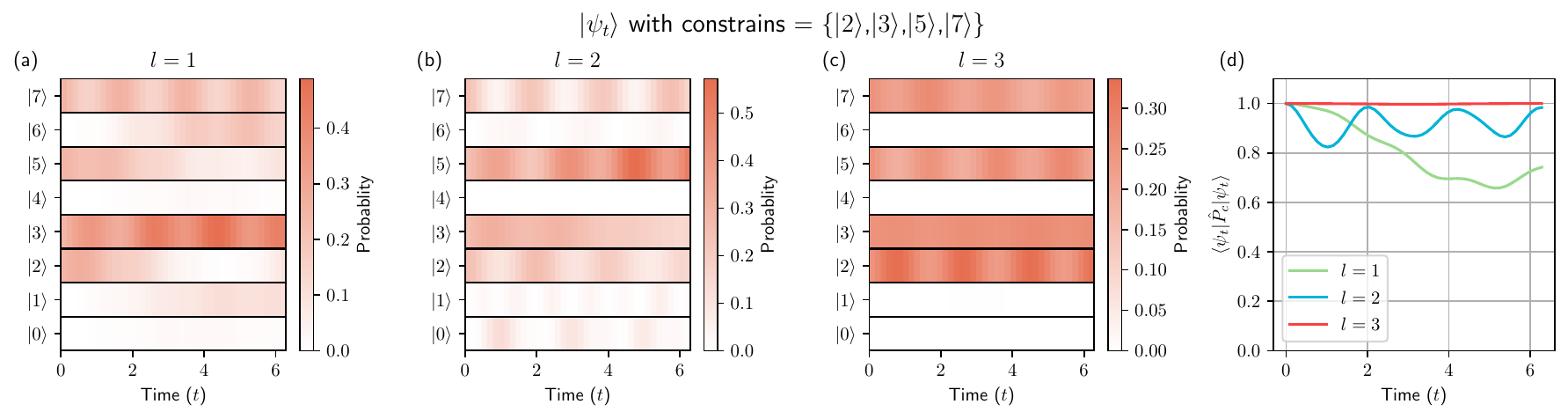}
  \caption{Probability $\abs{\bra{i}\exp{-iH_{qcl}t}\ket{\psi_0}}^2$ for various values of learning layers given by $l=$ (a) 1 (b) 2 (c) 3 for a 3 qubit system with constraints shown in the title. In (d) we plot the projection of the wave function at time $t$ on to the constrained subspace.}
  \label{fig:sm_quantum_leak_1}
\end{figure}

To understand this better, let us look at some numerical examples. In the training cycle, one could potentially train the unitary $\exp{-iH_{qcl}t}$ for $k$ different time steps $t_1,t_2,\ldots,t_k$. Instead, for these experiments (and the numerical experiments in the main paper), we choose $k=1$ and set $t_1=1$. In the tests we conducted, on a well converged cost, learning for just $k=1,t_1=1$ seems to extend to any general $t$. We will first look at the leakage at the level of learning. To this end, in \autoref{fig:sm_quantum_leak_1}, we plot the probability $\abs{\bra{i}\ket{\psi_t}}^2$ $\forall i \in \{0,1,\ldots2^n\}$ where $\ket{\psi_t}=\exp{-iH_{qcl}t}\ket{\psi_0}$, with $\ket{\psi_0}$ being the equal superposition of all constrained states. We choose a 3 qubit system as an example, with constraints arbitrarily set to $\{\ket{2},\ket{3},\ket{5},\ket{7}\}$. In the three subplots, we plot the probabilities which are evolved with a $l$ layer PQC. $l$ is directly proportional to the quality of QCL training and is inversely related to the training cost. Since we are training the unitary to have connections within the subspace, $\abs{\bra{i}\ket{\psi_t}}^2$ should ideally be zero for all $\ket{i}$ not in the constraint. As seen from figures (a) and (b), in the case of $l=1$, there is significant leakage of the wave function into the states that are not inside the constraint space. This is sharply reduced as one increases the training layer, thus lowering the final cost.

Another way to quantify the leakage is to look at the projection of the evolved wave function on to the constrained subspace. This can be done by using the constrained projection operator $\hat{P}_c$, given by
\begin{align}
  \hat{P}_c=\sum_{x_i\in H_c}\ket{x_i}\bra{x_i}\label{eq:sm11},
\end{align}
where $\ket{x_i}$ belongs to the constrained subspace. \autoref{fig:sm_quantum_leak_1}(d) plots $\expval{\hat{P}_c}{\psi_t}$ for various values of the PQC layer. In the absence of any quantum leakage, this expectation value must be equal to $1$ as the wave function is completely inside the closed constrained space, and thus any deviation away from $1$ is an indication of quantum leakage. 

\begin{figure}[t]
  \includegraphics[width=0.5\columnwidth]{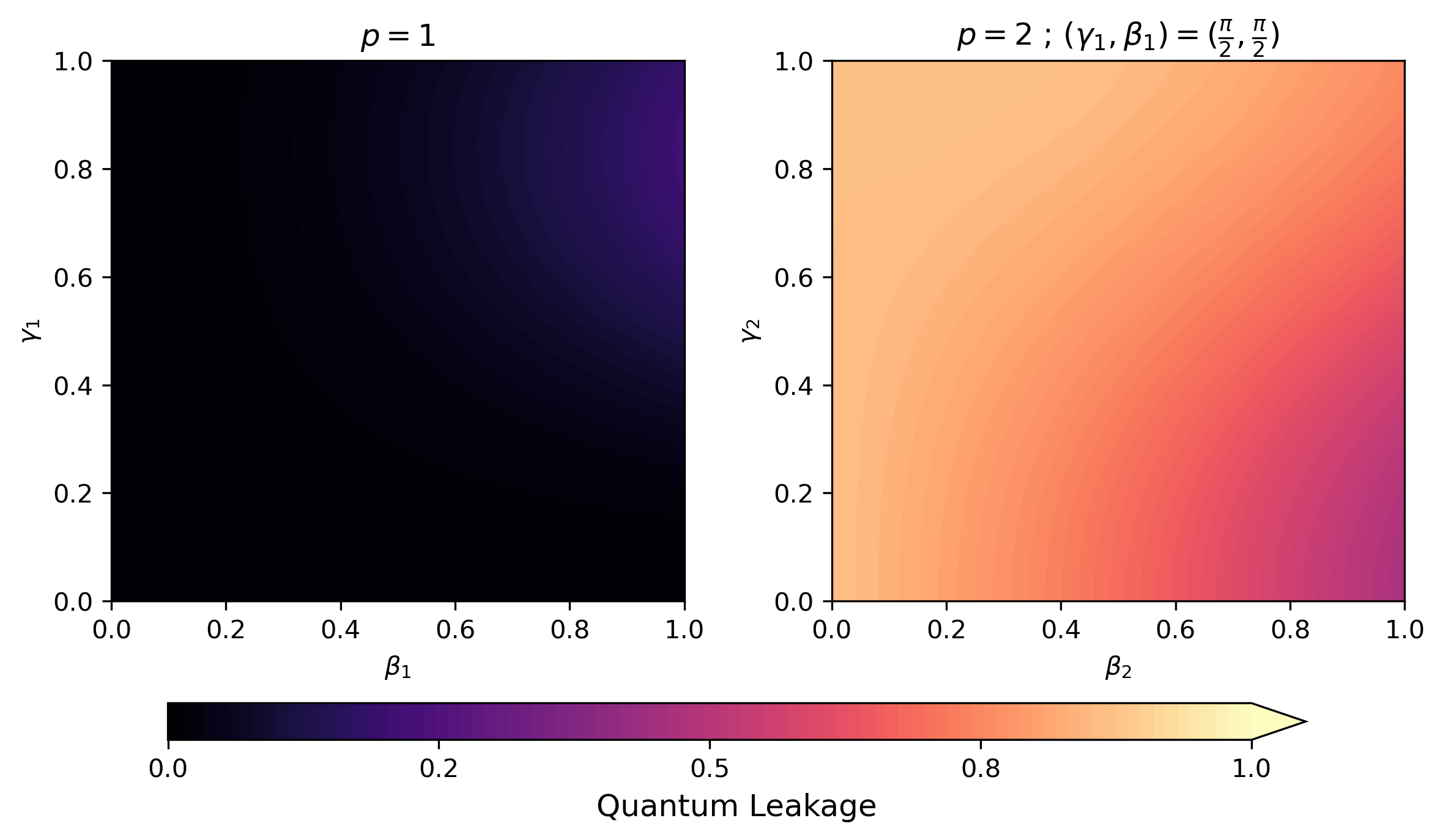}
  \caption{$\bra{\boldsymbol{\beta}, \boldsymbol{\gamma}} \mathds{1}-\hat{P}_c\ket{\boldsymbol{\beta}, \boldsymbol{\gamma}}$ for (a) $p=1$ (b) $p=2$ with first depth angles set to $\frac{\pi}{2}$.}
  \label{fig:sm_quantum_leak_2}
\end{figure}

Finally, using $\hat{P}_c$, we now study the effect of leakage on the combined QCL-QAOA routine. As mentioned in the main paper, the application of $\exp{-i H_{qcl}\gamma}$ is done at each level of the QAOA layer, and thus any small leakage is amplified as the QAOA depth is increased. Let $\ket{\boldsymbol{\beta}, \boldsymbol{\gamma}}_i$ denote the resulting wave function at the $i^{th}$ layer of the depth $p$ QAOA, this is equivalent to $\ket{\psi_1},\ket{\psi_2},\ldots\ket{\psi_i}$ shown in \autoref{fig:sm_algo}(b). We then plot the expectation value of $\mathds{1}-\hat{P}_c$ with respect to $\ket{\boldsymbol{\beta}, \boldsymbol{\gamma}}_i$ for $p=1$ (a) and $p=2$ (b) in \autoref{fig:sm_quantum_leak_2} as a color plot. Note that for $p=2$, we have 4 free variables, so we set $\gamma_1,\beta_1=\frac{\pi}{2}$ and plot as a function of$\gamma_2,\beta_2$. This quantity shows the leakage after applying both the cost unitary and the mixing unitary. Ideally, when one makes the final measurement of the QCL-QAOA cost, one expects that the final resulting wave function is inside the constrained Hilbert space. In the presence of leakage, leaked wave function results in sampling solutions that do not satisfy the constraint lead to low-quality results. This is encoded in quantity $\mathds{1}-\hat{P}_c$, and instead of looking at the final optimized QCL-QAOA ansatz, we explore the entire (or part) of the cost space. As is evident from the figure, the leakage ($\mathds{1}-\hat{P}_c$) is almost close to $0$, indicating there is little to no leakage at the level of the first mixer application. In (b), we see a drastic increase in leakage. This shows that one needs to converge the learning cost based on the use case of how deep the resulting QAOA is designed to be.

\begin{figure}
  \begin{subfigure}{0.4\columnwidth}
    \centering\includegraphics[width=7cm]{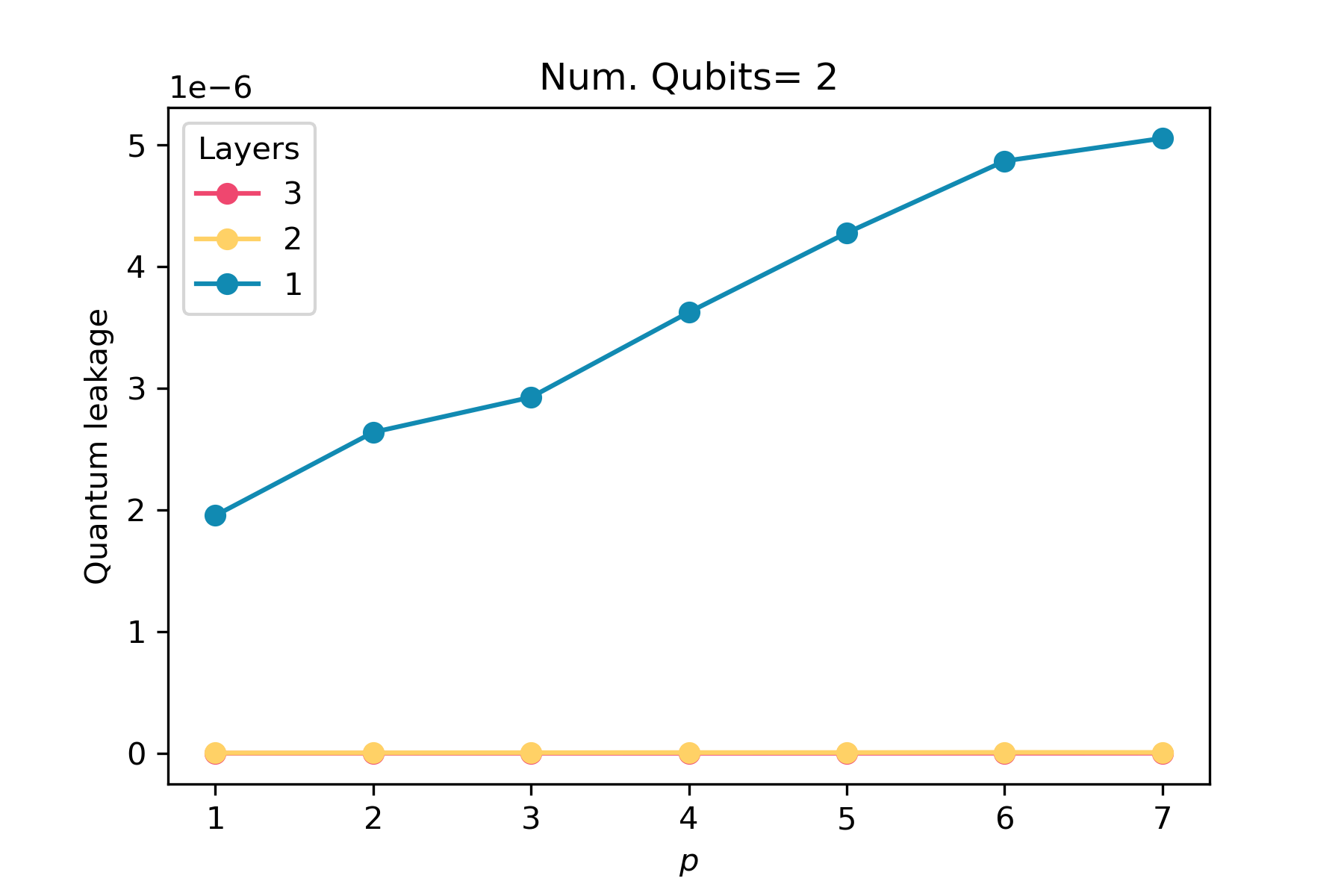}
    \caption{$2$ qubit system}
  \end{subfigure}
  \begin{subfigure}{0.4\columnwidth}
    \centering\includegraphics[width=7cm]{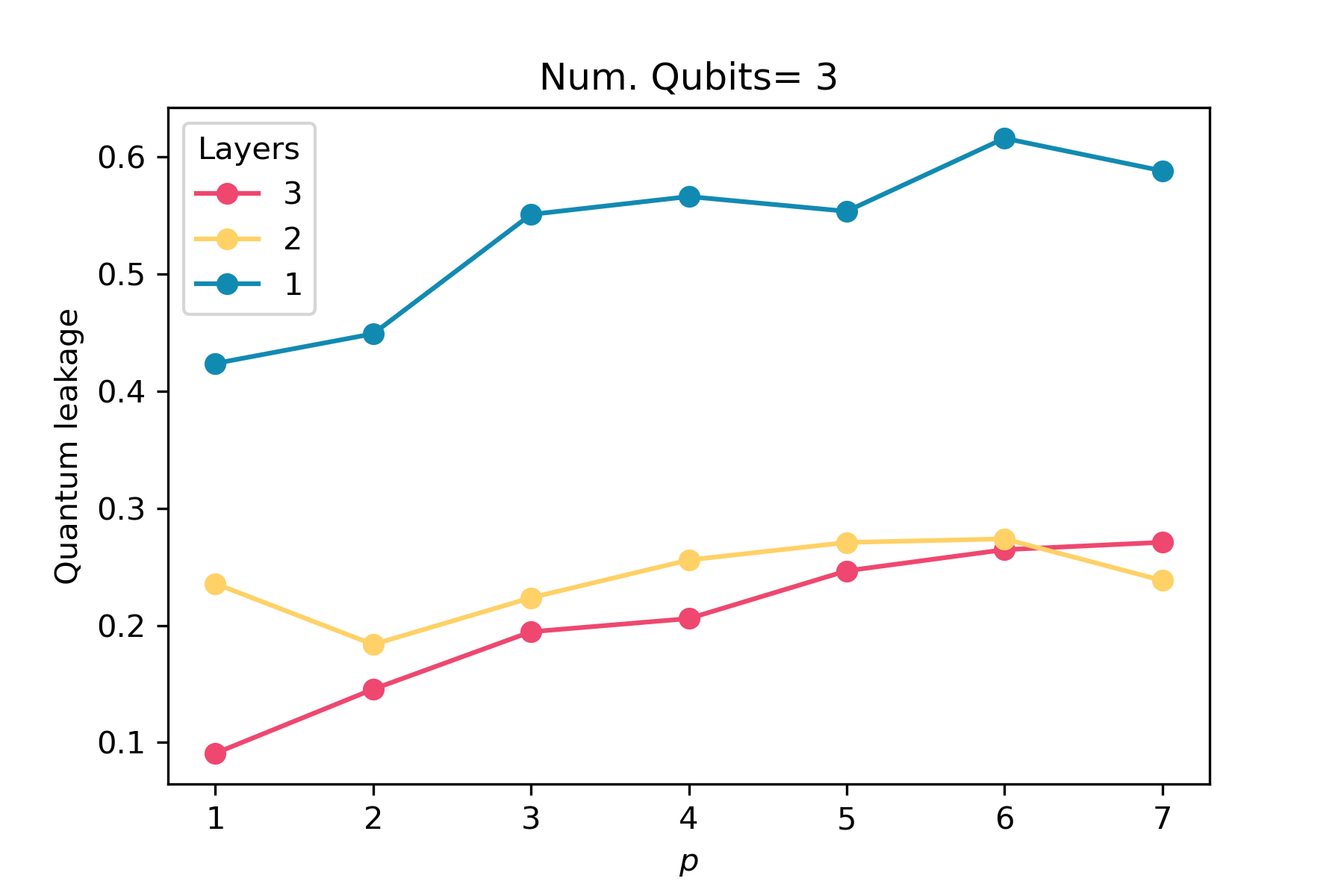}
    \caption{$3$ qubit system}
  \end{subfigure}
  \caption{Total quantum leakage for an instance of QCL-QAOA as function of $p$ for different layers. Note the difference in scale between the two ./figures}\label{fig:sm_total_leak}
\end{figure}

We investigate this further by looking at the total leakage at a given layer. We do this by computing the integral of the leakage throughout the volume defined by the $2p$ parameters which is given as $\int_{V_p}\expval{\mathds{1}-\hat{P}_c}{\boldsymbol{\beta}, \boldsymbol{\gamma}}$. We numerically do this for various values of $p$. Since integration techniques like  trapezoidal rule, Simpson's rule are costly at this scale, we use Monte-Carlo integration to compute the integral. \autoref{fig:sm_total_leak} shows the total quantum leakage for (a) two-qubit and (b) three-qubit systems as function of $p$. We also plot the same for various values of PQC layer depth. We see an almost linear trend with respect to $p$ for the total leakage. This shows that if one starts with a bad cost learning error, the leakage at worst gets linearly spilled to the next depth.

\section{Wasserstein performance metric $(\zeta)$}
In this section, we will explain  the performance metric given in the main text, which uses Wasserstein distance in detail. We first give a formal definition of  Wasserstein distance and then give a more intuitive practical overview of it and how it fits into the context of our performance metric.

Given two Borel probability distributions $p$ and $q$ over $\Sigma$ and $r\in[1,\infty)$, the $r-$Wasserstein distance $W_r(p,q)\in [0,\infty]$ between $p$ and $q$ is defined by \cite{cuesta1989notes}
\begin{align}
  W_{r}(p, q):=\inf _{\mu \in \Pi(p, q)}\left(\mathbb{E}_{(X, Y) \sim \mu}\left[\rho^{r}(X, Y)\right]\right)^{1 / r},\label{eq:sm12}
\end{align}
where $\Pi(p,q)$ denotes all couplings between $X \sim p$ and $Y \sim q$; that is,

\begin{align}
  \Pi(p,q) \coloneqq \{\mu:\Sigma^2 \rightarrow[0,1] \quad \forall \, A\in \Sigma,\, \mu(A\times \Omega)=p(A) \text{ and } \mu ( \Omega \times A)=q(A) \}\label{eq:sm13},
\end{align}
 is the set of joint probability measure $\mu$ over $\Omega \times \Omega $ with marginals $p$ and $q$. Intuitively, in the discrete case, $W_r(p,q)$ is the $r-$ weighted total cost of transforming a mass weighted distributed according to $p$ to be distributed according to $q$, where the cost of moving a unit mass from $x\in\Omega$ to $y\in\Omega$ is $\rho(x,y)$. \autoref{eq:sm12} generalizes this intuitive measure to arbitrary probability measures. It is important to note that $W_r$ is symmetric with respect to $p$ and $q$ and satisfies the triangle inequality with $W_r(p,p)=0$, and in the case where $\rho$ is (not)a metric, $W_r$ is a (pseudo)metric\cite{Clement_2008,Figalli_2010}. Despite its origins in optimal transport, $W_r$ is utilized in such diverse areas as probability theory, physics, statistics, economics, Machine learning\cite{Fournier_2014,Mohajerin_Esfahani_2017,Analui_2014} \textit{etc.}. The notion of $W_r$ has even been formally extended to pure quantum setting \cite{Agredo_2017,depalma2020quantum}.
 
For our use case of performance metric, the notion of $W_r$  is intuitively used as a distance measure between the found optimal state $\ket{\psi^*}$ and the ideal optimal distribution $i.e.$ a $\delta-$function on the optimal solution. To do this, one needs a supporting metric for the states. In our case, the QUBO cost value of our basis set provides a natural supporting metric. For instance, distance between $\ket{i}$ and $\ket{j}$ could be taken as $\abs{i-j}$ where $H_{cost}\ket{i}=i\ket{i}$. There are two major problems with this; first, this distance metric is unbounded, and second, there is no obvious way to extend this in case of constrained search. To this end, we attach an artificial metric to the basis set after mapping them to a new basis which essentially rearranges them. We map each basis $\ket{i}$  to new basis $\ket{\tilde{i}}$, which are first ordered by their ability to satisfy the constrain and then by their cost.

\begin{table}[htbp]
  
  \begin{ruledtabular}
  \begin{tabular}{c|cccccccc}
    Original basis & $\ket{0}$ & $\ket{1}$ & $\ket{2}$ & $\ket{3}$ & $\ket{4}$ & $\ket{5}$ & $\ket{6}$ & $\ket{7}$ \\ \hline
    Constraint & 1 & 1 & 0 & 0 & 1 & 1 & 0 & 0 \\ \hline
    QUBO Cost & 0.3 & 2 &3.6& 0.4 & 0.1 & 1.4 & 0.2 &  5.5 \\\hline
    Mapped basis & $\ket{\tilde{1}}$ & $\ket{\tilde{3}}$ & $\ket{\tilde{6}}$ & $\ket{\tilde{5}}$ & $\ket{\tilde{0}}$ & $\ket{\tilde{2}}$ & $\ket{\tilde{4}}$ & $\ket{\tilde{7}}$ \\ 
  \end{tabular}
  \end{ruledtabular}
  \caption{Example of mapping the original basis to the new basis and the corresponding QUBO cost along with the constrain satisfaction criteria ($1$ for satisfied and $0$ for not satisfied)}\label{tab:sm_mapping}
\end{table}

Let us look at a concrete example for a $3$ qubit system. \autoref{tab:sm_mapping} shows the QUBO cost associated with the basis bit strings denoted by $\ket{i}$. In the first column, we show the basis in our usual order where $\ket{i}$ corresponds to the $i^{th}$ binary bit string. The second and third columns show if $\ket{i}$ satisfies the constraint and the QUBO cost associated with the bit string, respectively. To get the mapped bit string, we first assign the states $\ket{\tilde{0}}$ to the smallest QUBO cost bit string that satisfies the constrain and go on assigning in ascending order for all bit strings that satisfy the constrain, after which we do the same for those that do not satisfy the constraint.

\begin{figure}[h]
  \includegraphics[width=1\columnwidth]{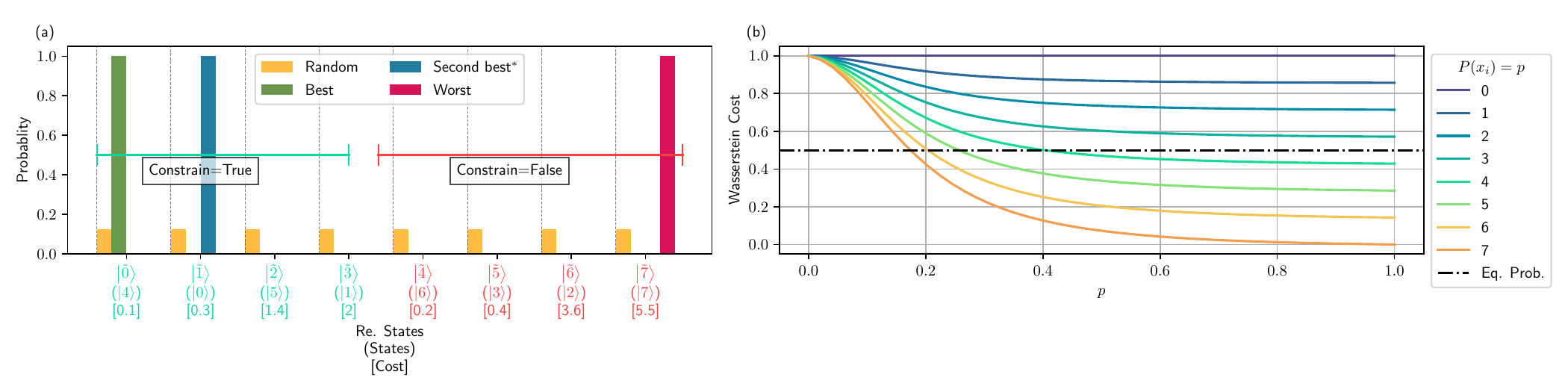}
  \caption{(a) Example of a QUBO problem and the corresponding mapping as given in \autoref{tab:sm_mapping}. We show the best, worst, second best$^*$ and random states. (b) Performance metric $\zeta$ for $\ket{\psi_i(p)}$ as defined in the text. }
  \label{fig:sm_cost}
\end{figure}

With this mapping, it is now trivial to see that a state given by $\ket{\psi}=\ket{\tilde{0}}$ is the best optimal solution while $\ket{\tilde{2^n}}$ is the worst for a $n-$ qubit system. In \autoref{fig:sm_cost}(a), we show the same, where the states are arranged now in the mapped basis. We also plot the state $\ket{\tilde{1}}$, the second-best quality solution ($^*$by best quality solution, we mean solutions where the probability is not spread out, but instead concentrated on a single state \textit{i.e.} $\delta$ function). As we proceed further down the plot, a $\delta$ function solution gets worse and worse until it reaches $\ket{\tilde{2^n}}$. Finally, since the QUBO cost depends on the problem, using the same as a metric for the states makes the performance metric unbounded. For this reason, we use a trivial metric, where the distance between adjacent mapped basis is given by $\frac{1}{2^n}$.  Thus, with the setup, we define our performance metric ($\zeta$) for a given optimal state $\ket{\psi^*}$ as 

\begin{align}
\zeta\left(\ket{\psi^*}\right)=1-W_1\left(\ket{\tilde{0}},\ket{\psi^*}\right)\label{eq:sm14},
\end{align}

where $W_1(\cdot,\cdot)$ is $1-$Wasserstein distance. To get an intuitive idea of the metric, in \autoref{fig:sm_cost}(b), we plot $\zeta$ for states given as $\ket{\psi_i(p)}=(1-p)\ket{\tilde{0}}+p\ket{\tilde{i}}$ with $p \in [0,1]$ for all $i\in[0,7]$. As one can see, this metric is bounded by $[0,1]$ with the best(worst) state being $1$($0$). For a random state, (\autoref{fig:sm_cost}(a) yellow), the performance metric turns out to be $0.5$. Any constraining algorithm needs to perform better than a random state generated from within the constrain space; thus, one can use this performance metric to identify how quantitatively good the solution is in constraining the solution to search space. Thus, when compared, the absolute value of metric indicates how well a given algorithm performs compared to a random algorithm having a value of 0.5. Note that this metric can be used for hard and soft constrain techniques.

\putbib

\end{bibunit}
\end{document}